\documentclass[10pt,twocolumn,twoside]{IEEEtran}
\usepackage{resizegather}
\usepackage{multirow}
\usepackage{bigstrut}
\usepackage{cite}

\usepackage{xcolor}
\usepackage{amsmath}
\usepackage{mathtools}
\usepackage{tikz}
\usepackage{verbatim}
\usepackage{subcaption}
\usepackage{pgfplots}
\usepackage{soul}
\pgfplotsset{compat=1.10}
\usepackage{graphicx}
\usepackage{epstopdf}
\usepackage{amssymb}
\usepackage{relsize}
\usepackage[textwidth=3.8em,textsize=scriptsize,disable]{todonotes}
\usepackage{algorithm}
\definecolor{Gray}{gray}{0.90}
\definecolor{blue}{rgb}{0,0,0}
\usepackage{colortbl}

\DeclareMathOperator*{\argmin}{arg\,min}
\usepackage{pgfplots}
\usepackage{algorithm}
\usepackage{algorithmicx}
\usepackage{algcompatible}
\usepackage{algpseudocode}
\newtheorem{theorem}{\bf{Theorem}}[section]

\newtheorem{cor}[theorem]{\bf{Corollary}}
\newtheorem{lem}[theorem]{\bf{Lemma}}
\newtheorem{prop}[theorem]{\bf{Proposition}}

\newtheorem{observation}[theorem]{Observation}
\newtheorem{fact}[theorem]{Fact}
\newtheorem{remark}[theorem]{\bf{Remark}}
\newenvironment{definition}[1][Definition]{\begin{trivlist}
\item[\hskip \labelsep {\bfseries #1}]}{\end{trivlist}}
\def\qed{\hfill\rule[-1pt]{5pt}{5pt}\par\medskip}
\usepackage{amssymb}
\usepackage{url}

\newcommand{\calX}[0]{\mathcal{X}}

\newcommand{\calN}[0]{\mathcal{N}}

\newcommand{\calD}[0]{\mathcal{D}}

\newcommand{\calO}[0]{\mathcal{O}}
\title{\LARGE \bf
Computation of the Distance-based Bound on Strong Structural Controllability in Networks
}
\author{ Mudassir~Shabbir,
         Waseem~Abbas,~\IEEEmembership{Member,~IEEE},
        A. Yasin Yaz{\i}c{\i}o\u{g}lu,~\IEEEmembership{Member,~IEEE},\\
        Xenofon Koutsoukos,~\IEEEmembership{Fellow,~IEEE}
\thanks{ M.~Shabbir is with the Electrical Engineering and Computer Science at the Vanderbilt University, Nashville, TN, USA. E-mail: \texttt{mudassir@vanderbilt.edu}. W.~Abbas is with the Department of Systems Engineering at the University of Texas at Dallas, Richardson, TX, USA. E-mail: \texttt{waseem.abbas@utdallas.edu}. X.~Koutsoukos is with the Electrical Engineering and Computer Science Department at the Vanderbilt University, Nashville, TN, USA. E-mail: \texttt{xenofon.koutsoukos@vanderbilt.edu}. A. Y.~Yaz{\i}c{\i}o\u{g}lu is with the Department of Electrical and Computer Engineering at the University of Minnesota, Minneapolis, MN, USA. E-mail: \texttt{ayasin@umn.edu}}

\thanks{Some preliminary results appeared in \cite{Mudassir2019}.}

}

\IEEEoverridecommandlockouts
\begin{document}

\maketitle

\begin{abstract}
In this paper, we study the problem of computing a tight lower bound on the dimension of the strong structurally controllable subspace (SSCS) in networks with Laplacian dynamics. The bound is based on a sequence of vectors containing the distances between leaders (nodes with external inputs) and followers (remaining nodes) in the underlying network graph. Such vectors are referred to as the distance-to-leaders vectors. {We give exact and approximate algorithms to compute the longest sequences of distance-to-leaders vectors, which directly provide distance-based bounds on the dimension of SSCS. The distance-based bound is known to outperform the other known bounds (for instance, based on zero-forcing sets), especially when the network is partially strong structurally controllable. Using these results, we discuss an application of the distance-based bound in solving the leader selection problem for strong structural controllability. Further, we characterize strong structural controllability in path and cycle graphs with a given set of leader nodes using sequences of distance-to-leaders vectors. Finally, we numerically evaluate our results on various graphs.} 

\end{abstract}
\section{Introduction}
\label{sec:introduction}
Network controllability has been an important research topic in network science and control. The notion of strong structural controllability accounts for the controllability of all such networks that have the same \emph{structure} of an underlying network graph, that is, networks having the same vertex and edge sets but possibly different (non-zero) edge weights. A network is strong structurally controllable (SSC) with a given set of input (leader) nodes if it is controllable for any choice of (non-zero) edge weights in the underlying network graph. There exist efficient algorithms to verify the strong structural controllability of networks \cite{chapman2013strong,jarczyk2011strong,weber2014linear,monshizadeh2014zero}. If a network is not SSC with a given set of leader nodes, it is of interest to determine how far the network is from becoming SSC, or roughly speaking, how much of the network is always controllable. More formally, this issue is concerned with computing the dimension of the \emph{strong structurally controllable subspace (SSCS)} (defined in Section~\ref{sec:SSC}), which is related to an NP-hard problem of finding the minimum rank of a structure (or pattern) matrix \cite{fazel2004rank,fallat2007minimum,Bhangale2015complexity}.

In this paper, we study the problem of computing a tight lower bound on the dimension of SSCS of networks with Laplacian dynamics. Since the exact computation of is challenging, various bounds have been proposed in the literature \cite{monshizadeh2014zero,zhang2014upper,yaziciouglu2016graph,van2017distance,mousavi2018structural}. Here, we consider a tight lower bound proposed in \cite{yaziciouglu2016graph}, which relates the notion of strong structural controllability to the {distances} between nodes in the underlying network graph. In \cite{YasinCDC2020}, we compare this \emph{distance-based} bound with another widely used bound based on the notion of zero-forcing sets \cite{monshizadeh2014zero,monshizadeh2015strong,mousavi2016controllability,mousavi2018structural}. Our analysis in \cite{YasinCDC2020} shows that the distance-based bound is typically better than the zero-forcing-based bound, especially when the network is not completely strong structurally controllable. \textcolor{blue}{Additionally, the distance-based bound can be applied in exploring the trade-off between controllability and robustness in networks with Laplacian dynamics \cite{AbbasACC2019}, edge augmentation in networks while preserving their strong structural controllability \cite{AbbasACC2020Augmentation} and designing a leader selection algorithm \cite{yaziciouglu2016graph}. It also has applications for target controllability in linear networks, where
the goal is to control a subset of agents (targets) instead of the entire network by injecting input through leader nodes \cite{van2017distance,monshizadeh2015strong}.} Despite advantages, efficient computation of the distance-based bound has been an issue, especially in large networks.

To compute the distance-based bound on the dimension of SSCS, an algorithm has been presented in \cite{yaziciouglu2016graph} that takes $O(m^n)$ time, where $n$ is the total number of nodes in the network and $m$ is the number of leader nodes. Here, we present an algorithm that takes $O(m(n\log n + n^m))$ time to compute the distance-based bound, which is a significant improvement. We note that for a fixed number of leaders, the algorithm is polynomial in the number of nodes. For instance, in the case of two leaders, our algorithm takes $O(n^2)$ time as compared to the $O(2^n)$ runtime of the algorithm in \cite{yaziciouglu2016graph}. When the number of leaders is on the order of $n$, the algorithm will take exponential time. For such cases, we also present a \emph{greedy} algorithm that approximates the distance-based bound and runs in $O(m n\log n)$ time. In our experiments, we observe that the bound returned by the greedy algorithm is very close to the optimal in almost all cases.

The main idea of the distance-based bound is to obtain distances between leaders and other nodes, arrange them in vectors called \emph{distance-to-leaders vectors}, and then construct a sequence of such vectors, called as \emph{Pseudo-monotonically increasing (PMI) sequence}, which satisfies some monotonicity conditions (as explained in Section \ref{sec:PMI_bound}). Computing distances between nodes is straightforward; however, constructing an appropriate PMI sequence, whose length provides a bound on the dimension of SSCS, is computationally challenging. We provide efficient algorithms with performance guarantees to compute such sequences. 

\emph{Contributions:} We provide dynamic programming-based exact algorithm that runs in $O(m(n\log n + n^m))$ time to compute an optimal PMI sequence of distance-to-leaders vectors consisting of distances between leaders and other nodes. Here $m$ and $n$ denotes the number of leaders and nodes, respectively. The length of the sequence directly gives a tight lower bound on the dimension of SSCS of networks with Laplacian dynamics. We also propose an approximation algorithm that computes a near-optimal PMI sequence of distance-to-leaders vectors in practice and takes $O(m n\log n)$ time. If there exists a PMI sequence of distance-to-leaders vectors of length $n$, then the network is strong structurally controllable and the greedy algorithm always returns such a sequence. Further, We analyze PMI sequences of distance-to-leaders vectors in paths and cycles with arbitrary leaders. We also discuss the application of distance-based bound in solving the leader selection problem for strong structural controllability and also provide a numerical evaluation of results.

\subsection{Related Work}
\label{sec:Related_Work}
The notion of strong structural controllability was introduced in \cite{mayeda1979strong}, and the first graph-theoretic condition for single-input systems was presented. For multi-input systems, \cite{reinschke1992strong}~provided a condition to check strong structural controllability in $\calO(n^3)$ time, where $n$ is the number of nodes. Authors in  \cite{jarczyk2011strong} refined previous results and further provided a characterization of strong structural controllability. An algorithm based on constrained matching in bipartite graphs with time complexity $\calO(n^2)$ was given in \cite{chapman2013strong} to check if the system is strong structurally controllable with given inputs. In \cite{weber2014linear}, an algorithm with a runtime linear in the number of nodes and edges was presented to verify whether a system is strong structurally controllable. The relationship of strong structural controllability and zero forcing sets (ZFS) was explored in \cite{monshizadeh2014zero,trefois2015zero}, and it was established that checking if a system is strong structurally controllable with given input nodes is equivalent to checking if the set of input nodes is a ZFS in the underlying graph.

If a network is not strong structurally controllable, then the notion of strong structurally controllable subspace (SSCS) \cite{monshizadeh2015strong,van2017distance}, which is an extension of the ordinary controllable subspace, is particularly useful to quantify controllability. In fact, the dimension of such a subspace (formally described in Section \ref{sec:SSC}) quantifies how much of the network is controllable in the strong structural sense (that is, independently of edge weights) with a given set of inputs. Some lower bounds on the dimension of SSCS have been proposed in the literature. In \cite{monshizadeh2015strong}, a lower bound based on the derived set of input nodes (leaders) was presented, which was further studied in \cite{mousavi2016controllability,mousavi2018structural}. With a single input node, the dimension of SSCS can be at least the diameter of the underlying network graph \cite{zhang2014upper}. In \cite{yaziciouglu2016graph}, a tight lower bound on the dimension of SSCS was proposed that was based on the distances between leaders and other nodes in the graph. 
The bound was used to explore the trade-off between strong structural controllability and network robustness in \cite{AbbasACC2019}. 
Further studies in this direction include enumerating and counting strong structurally controllable graphs for a given set of network parameters (leaders and nodes) \cite{o2016conjecture,menara2017number}, leader selection to achieve desired structural controllability (e.g., \cite{yaziciouglu2013leader,olshevsky2015minimum,fitch2016optimal,tzoumas2016minimal,commault2013input,clark2017submodularity,pequito2015complexity,commault2019functional,li2019target}), and network topology design for a desired control performance (e.g., \cite{aguilar2015graph,pequito2017robust,mousavi2017robust,xue2017input,commault2019structural}).

The rest of the paper is organized as follows: Section~\ref{sec:Prelim} introduces notations and preliminary concepts. Section~\ref{sec:exact_algorithm} provides dynamic programming-based exact algorithm to compute a distance-based bound on the dimension of SSCS. Section~\ref{sec:comp_greedy} presents and analyzes a greedy approximation algorithm. Section~\ref{sec:Application} discusses application of the bound to the leader selection problem. Section \ref{sec:PathCycle} discusses cases of path and cycle graphs. Section~\ref{sec:eval} provides a numerical evaluation of results, and Section~\ref{sec:conclusion} concludes the paper.
\section{Preliminaries}
\label{sec:Prelim}

We consider a network of $n$ dynamical agents represented by a simple (loop-free) undirected graph $G=(V,E)$ where the node set $V = \{v_1,v_2,\ldots,v_n \}$ represents agents, and the edge set $E$ represents interconnections between agents.\footnote{The results presented can be extended to directed networks in a straightforward manner as the distance-based bound on the dimension of SSCS holds true for directed networks also \cite[Remark 3.1]{yaziciouglu2016graph}.} An edge between $v_i,v_j\in V$ is denoted by $e_{ij}$. The \emph{neighborhood} of node $v_i$ is ${\calN_i\triangleq\{v_j\in V: e_{ij} \in E\}}$. The \emph{distance} between $v_i$ and $v_j$, denoted by $d(v_i,v_j)$, is simply the number of edges on the shortest path between $v_i$ and $v_j$. $\mathbb{R}^+$ is the set of positive real numbers. The \emph{weight function} 
\begin{equation}
\label{eq:weight_fn}
w:E\rightarrow \mathbb{R}^+
\end{equation}

assigns positive weight $w(e_{ij})$ to the edge $e_{ij}$. These weights define the coupling strength between nodes.

Each agent $v_i\in V$ has a state $x_i(t)\in\mathbb{R}$ at time $t$ and the overall state of the system is ${x(t) = \left[\begin{array}{cccc}x_1(t) & x_2(t) & \cdots & x_n(t)\end{array}\right]^T\in\mathbb{R}^n}$. 
The agents update states following the Laplacian dynamics, 
\begin{equation}
\label{eq:dynamics}
\dot{x}(t) = -L_wx(t) + B u(t),
\end{equation}
where $L_w\in\mathbb{R}^{n\times n}$ is the weighted \emph{Laplacian matrix} of $G$ and is defined as $L_w = \Delta - A_w$. Here, $A_w\in\mathbb{R}^{n\times n}$ is the weighted \emph{adjacency matrix} defined as

\begin{equation}
\label{eq:A_matrix}
\left[ A_w \right]_{ij} =
\left\lbrace
\begin{array}{ccc}
w(e_{ij}) & \text{if } e_{ij}\in E,\\
0 	   & \text{otherwise,}
\end{array}
\right.
\end{equation}
and $\Delta\in\mathbb{R}^{n\times n}$ is the \textit{degree matrix} whose entries are \textcolor{blue}{
\begin{equation}
\label{eq:Delta_matrix}
\left[ \Delta \right]_{ij} =
\left\lbrace
\begin{array}{ccc}
\sum_{k = 1}^{n} \left[ A_w \right]_{ik}& \text{if } i=j\\
0 	   & \text{otherwise.}
\end{array}
\right.
\end{equation}
}The matrix $B\in\mathbb{R}^{n\times m}$ in \eqref{eq:dynamics} is an \textit{input matrix}, where $m$ is the number of  leaders (inputs), which are the nodes to which external control signals are applied. Let ${V_\ell = \{\ell_1,\ell_2,\cdots,\ell_m\}\subseteq V}$
be the set of \emph{leaders}, then
\begin{equation}
\label{eq:Delta_matrix}
\left[ B \right]_{ij} =
\left\lbrace
\begin{array}{ccc}
1& \text{if } v_i = \ell_j\\
0 	   & \text{otherwise.}
\end{array}
\right.
\end{equation}

\subsection{Strong Structural Controllability}
\label{sec:SSC}
\begin{figure}[htb]
\centering
\includegraphics[scale=0.65]{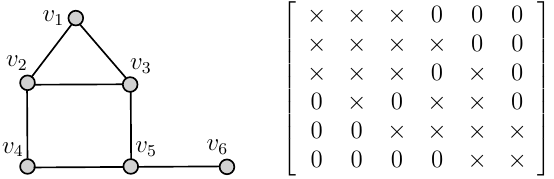}
\caption{A graph and its structured Laplacian matrix.}
\label{fig:Lap}
\end{figure}

A state $x_f\in\mathbb{R}^n$ is a \emph{reachable state} if there exists an input $u$ that can drive the network in \eqref{eq:dynamics} from \textcolor{blue}{any initial state $x_i$} to $x_f$ in a finite amount of time. A network $G=(V,E)$ in which edges are assigned weights according to the weight function $w$ in \eqref{eq:weight_fn}, and contains $V_\ell\subseteq V$ leaders is called \emph{completely controllable} if every point in $\mathbb{R}^n$ is reachable. Complete controllability can be checked by computing the rank of the \emph{controllability matrix}, $
\small
\Gamma(L_w,B) =
\left[
\begin{array}{ccccc}
B & (-L_w)B & (-L_w)^2B & \cdots & (-L_w)^{n-1}B
\end{array}
\right]
$. The network is completely controllable if and only if the rank of $\Gamma(L_w,B)$ is $n$, and in such case $(L_w,B)$ is called a \emph{controllable pair}. \textcolor{blue}{The range space of $\Gamma(L_w,B)$ describes the set of all reachable states, also called the controllable subspace. Thus, the rank of $\Gamma(L_w,B)$ is the dimesnion of the controllable subspace.} Note that edges in $G$ define the \emph{structure}---location of zero and non-zero entries in the Laplacian matrix---of the underlying graph, for instance, see Figure~\ref{fig:Lap}. The rank of resulting controllability matrix depends on the weights assigned to edges. For a given graph $G=(V,E)$ and leaders~$V_\ell$, rank$(\Gamma(L_w,B))$ could be different from $\text{rank}(\Gamma(L_{w'},B))$, where $w$ and $w'$ are two different choices of weight functions.

A network $G=(V,E)$ with $V_\ell$ leaders is \emph{strong structurally controllable} if and only if $(L_w,B)$ is a controllable pair for any choice of weight function $w$, or in other words, rank$(\Gamma(L_w,B))=n$ for all weight functions $w$. At the same time, the \emph{dimension of strong structurally controllable subspace (SSCS)}, denoted by $\gamma(G,V_\ell)$, is

\begin{equation}
\label{eq:dim_SSC}
\gamma(G,V_\ell) = \min\limits_{w} \left(\text{rank}\;\Gamma(L_w,B)\right).
\end{equation}
The minimum is taken over all weight functions $w$ in \eqref{eq:weight_fn}. \textcolor{blue}{Thus, $\gamma(G,V_\ell)$ is the minimum dimension of the controllable subspace that can be attained from $G$ with $V_\ell$ leaders and any choice of feasible edge weights.}


\subsection{Distance-based Lower Bound on the dimension of SSCS}
\label{sec:PMI_bound}
We use a tight lower bound on the dimension of SSCS as proposed in \cite{yaziciouglu2016graph}. The bound is based on the distances between nodes in a graph. Assuming $m$ leaders $V_\ell = \{\ell_1,\cdots,\ell_m\}$, we define a vector of non-negative integers called as the \emph{distance-to-leaders} vector for a node $v_i\in V$ as

\begin{equation*}
\label{eq:DLvector}
D_i = \left[
\begin{array}{ccccc}
d(\ell_1,v_i) & d(\ell_2,v_i)  & \cdots & d(\ell_m,v_i)
\end{array}
\right]^T.
\end{equation*}
\textcolor{blue}{The $j^{th}$ component of $D_i$, denoted by $[D_i]_j$, is $ d(\ell_j,v_i)$, the distance between leader $\ell_j$ and the node $v_i$}. Next, we define a sequence of distance-to-leaders vectors, called as \emph{pseudo-monotonically increasing} sequence below.

\begin{definition}{(\emph{Pseudo-monotonically Increasing (PMI) Sequence)}}
\label{def:PMI_def}
\label{def:PMI}
Let $\calD$ be a sequence of distance-to-leaders vectors and $\calD_i$ be the $i^{th}$ vector in the sequence. We denote the $j^{th}$ component of the vector $\calD_i$  by $[\calD_i]_j$. \textcolor{blue}{Then, $\calD$ is PMI if for every $\calD_i$ in the sequence, there exists some $\pi(i)\in\{1,2,\cdots, m\}$ such that} 

\textcolor{blue}{
\begin{equation}
\label{eq:PMIcondition}
[\calD_i]_{\pi(i)} < [\calD_{j}]_{\pi(i)}, \;\;\forall j > i,
\end{equation}
}
\textcolor{blue}{i.e., the above condition needs to be satisfied for all the subsequent distance-to-leader vectors $\mathcal{D}_j$ appearing after $\mathcal{D}_i$ in the sequence.} 
Here, $m$ is the number of leaders. \textcolor{blue}{We say that $\calD_i$ satisfies the \emph{PMI property} at coordinate $\pi(i)$ whenever $[\calD_i]_{\pi(i)} < [\calD_{j}]_\pi(i),\;\forall j>i$.}
\end{definition}

\begin{figure}[htb]
\centering
\includegraphics[scale=0.8]{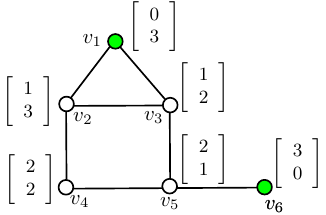}
\caption{A network with two leaders $V_\ell = \{\ell_1,\ell_2\} = \{v_1,v_6\}$, along with the distance-to-leaders vectors of nodes. A PMI sequence of length five is $\calD = [\calD_1 \; \calD_2 \; \cdots \; \calD_5] = [D_6 \; D_5 \; D_1 \; D_4 \; D_2]$.}
\label{fig:PMI}
\end{figure}

An example of distance-to-leaders vectors is illustrated in Figure \ref{fig:PMI}. A PMI sequence of length five is

\begin{equation}
\label{eq:PMIexample}
\calD =
\left[
\left[
\begin{array}{c}
3\\ \textcircled{0}
\end{array}
\right]
,
\left[
\begin{array}{c}
2 \\ \textcircled{1}
\end{array}
\right]
,
\left[
\begin{array}{c}
\textcircled{0} \\ 3
\end{array}
\right]
,
\left[
\begin{array}{c}
2\\ \textcircled{2}
\end{array}
\right]
,
\left[
\begin{array}{c}
\textcircled{1} \\ 3
\end{array}
\right]
\right].
\end{equation}

Indices of circled values in \eqref{eq:PMIexample} are the coordinates at which the corresponding distance-to-leaders vectors are satisfying the PMI property.
The length of the longest PMI sequence of distance-to-leaders vectors is related to the dimension of SSCS as stated in the following result.
\begin{theorem} \cite{yaziciouglu2016graph}
\label{thm:PMI}
If $\delta(G,V_\ell)$ is the length of the longest PMI sequence of distance-to-leaders vectors in a network $G=(V,E)$ with leaders $V_\ell$, then
\begin{equation}
\label{eq:PMI_bound}
\delta(G,V_\ell) \le \gamma(G,V_\ell).
\end{equation}
\end{theorem}

\textcolor{blue}{We note that for a given graph $G=(V,E)$ and leader nodes $V_\ell\subseteq V$, the length of the longest PMI sequence describes the minimum dimension of the controllable subspace for any feasible edge weights. In other words, if the length of the longest PMI is $k\le n$, then the dimension of the controllable subspace of the system is at least $k$, regardless of the edge weights. Moreover, the bound in \eqref{eq:PMI_bound} is tight,}
as discussed in \cite{yaziciouglu2016graph}. For instance, for path graphs in which one of the end nodes is a leader, and for cycle graphs in which two adjacent nodes are leaders, we have $\delta(G,V_\ell) = \gamma(G,V_\ell)$.
the dimension of SSCS and the length of the longest PMI sequence of distance-to-leaders vectors are equal, and hence $\delta(G,V_\ell) = \gamma(G,V_\ell)$. 
We discuss the length of the longest PMI sequences of distance-to-leaders vectors in path and cycle graphs with arbitrary leaders in Section \ref{sec:PathCycle}. 

Our main goal is to compute a PMI sequence of maximum length, and consequently, a lower bound on the dimension of SSCS. We provide an exact algorithm in Section \ref{sec:exact_algorithm} and a greedy approximation algorithm in Section \ref{sec:comp_greedy}. 

\section{Exact Algorithm for the Distance Bound}
\label{sec:exact_algorithm}
In this section, we provide a dynamic programming-based exact algorithm to compute a longest PMI sequence of distance-to-leaders vectors and, as a result, a distance-based lower bound on the dimension of SSCS. 

We note that each distance-to-leaders vector $D_i$ can be viewed as a point in $\mathbb{Z}^m$, and without loss of generality, we may assume that points $D_i$ are \emph{distinct}. Otherwise, we can throw away multiple copies of the same point since duplicate points can not satisfy the PMI property on any coordinate. The following observation is crucial to our algorithms.

\begin{observation}
	\label{obs1}
	Given a set of points $D_1,D_2,\ldots,D_n$, if there exists a point $D_i$ and an index $j$ such that $[D_i]_j<[D_{i'}]_j$ for all $D_i\neq D_{i'}$, then $D_i$ is a unique minimum point in the direction (coordinate) $j$ and there is a longest PMI sequence which starts with $D_i$. However, it is possible that there is no unique minimum in any direction. This leads us to the definition of a {\em conflict} and {\em conflict-partition}.
	\end{observation}
	
\begin{figure}[h!]
\centering
	\includegraphics[scale=0.85]{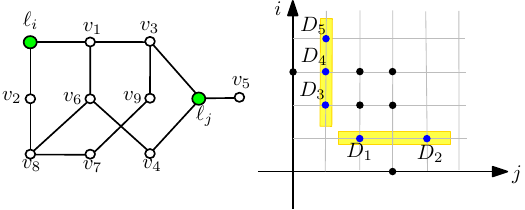}
	\caption{A graph with two leaders and a plot of distance-to-leaders vectors as points in a plane. Point set $\calX = \{D_1,D_2,D_3,D_4,D_5\}$ constitutes a conflict, where $\calX_i = \{D_3,D_4,D_5\}$ and $\calX_j=\{D_1,D_2\}$.}
	\label{fig:obs1}
\end{figure}

\begin{definition}{\em (Conflict-partition)}
\label{def:conflict}
A \textit{conflict} is a set of points $\calX$ that can be partitioned into $\calX_1,\calX_2,\ldots, \calX_m$ such that all points $D_p\in \calX_j$ have $[D_{p}]_j=[D_{q}]_j$ if $D_q\in \calX_j$, and $[D_{p}]_j\le [D_{q}]_j$ if $D_q\notin \calX_j$. Further, $|\calX_j|>1$ for all $j$. Such a partition is called {\em conflict-partition} or \textit{c-partition} for short.\footnote{In general, parts of a partition do not intersect. For the lack of a better term, we are slightly abusing this term in the sense that parts ($\calX_i$) intersect at most one element.}
\end{definition}
An example of conflict is illustrated in Figure~\ref{fig:obs1}.

%
%
%
%


It is easy to see that a PMI sequence
can not contain all points in a conflict. In fact, we can strictly bound the number of points from a conflict that can be included in a PMI sequence.
\begin{lem}
\label{obs:conflict_bound}
Let $\calX_1,\calX_2,\ldots, \calX_m$ be a c-partition of a conflict $\calX$ for a given set of points. Then any PMI sequence contains at most
$|\calX|-\min(|\calX_1|,|\calX_2|,\ldots,|\calX_m|)+1$ points from $\calX$.
\end{lem}

\textit{Proof:}
Let $k_j = |\calX_j|$ for all $1\le j\le m$ for the partition defined in the statement, then for the sake of contradiction, let's assume that there is a sequence $\calD'$ that contains more points from $\calX$. Let $D_p\in \calX_j$ be a point that appears first in $\calD'$. If $D_p$ satisfies PMI property on $j^{th}$ coordinate then the remaining $k_j-1$ points with the same minimum $j^{th}$ coordinate in $\calX_j$ can not be included in $\calD'$. So $D_p$ must satisfy PMI property on some ${j'}^{th}$ coordinate for the following points in the sequence, where ${j'}\ne j$. But then $\calD'$ must miss at least $k_{{j'}}$ points that have smaller or equal ${j'}$ coordinate by the definition of conflict, which is a contradiction. Thus, the claim follows.
\qed

As an example, consider a set of points $\calX = \{D_1,D_2,D_3,D_4,D_5\}$ in Figure~\ref{fig:obs1}. There are two points with the minimum $j^{th}$ coordinate and three points with the minimum ${j'}^{th}$ coordinate (where $j' = i$). If $D_1$ is picked as first point (among this set), we must either drop $D_2$ or all $D_3,D_4,D_5$ for future consideration in the PMI sequence. Similarly if $D_3$ is picked before everyone else, we can not pick either of $D_4,D_5$, or any of $D_1,D_2$
for future consideration regardless of the other points. Note that the bound in Lemma~\ref{obs:conflict_bound} is tight: if we remove $|\calX_j|-1$ points from the smallest part of a c-partition, all remaining points can satisfy the PMI property on coordinate $j$ unless some of these remaining points are included in any other conflict.

In the following, we use the following notations:
\begin{itemize}
\item $\L^j$ denotes a list of points ordered by the
non-decreasing $j^{th}$ coordinate. 
\item $\L^j_i$ denotes the $i^{th}$ point in the list $\L^j$, and
\item $\L^j_{i,k}$ is the (integer) value of $k^{th}$ coordinate of $\L^j_i$.\footnote{We recommend to use linked priority queues or similar data structure for these lists so that one could easily
delete a point from lists while maintaining respective orders in logarithmic time.}
\end{itemize}

Let $D$ be a set of $n$ points in $\mathbb{Z}^m$. We can sort all
points with respect to all coordinates beforehand, so our algorithm will get $m$ lists $\{\L^1,\L^2,\ldots,\L^m\}$ of $n$ points each as input. Next, we design an algorithm that is based on dynamic programming to compute the lower bound $\delta(G,V_\ell)$ in polynomial runtime when the number of leaders is fixed. Let $\{c_1,c_2,\ldots,c_m\}$ be a set of non-negative integers and $\calD^{[c_1,c_2,\ldots,c_m]}$ be a longest PMI sequence in which the value at $j^{th}$ coordinate of any point is at least $c_j$. Let $\alpha^{[c_1,c_2,\ldots,c_m]}$ be the length of such a sequence. Our algorithm will memoize on $\alpha^{[c_1,c_2,\ldots,c_m]}$.

\begin{figure}[h!]

	\centering
	\includegraphics[scale=0.53]{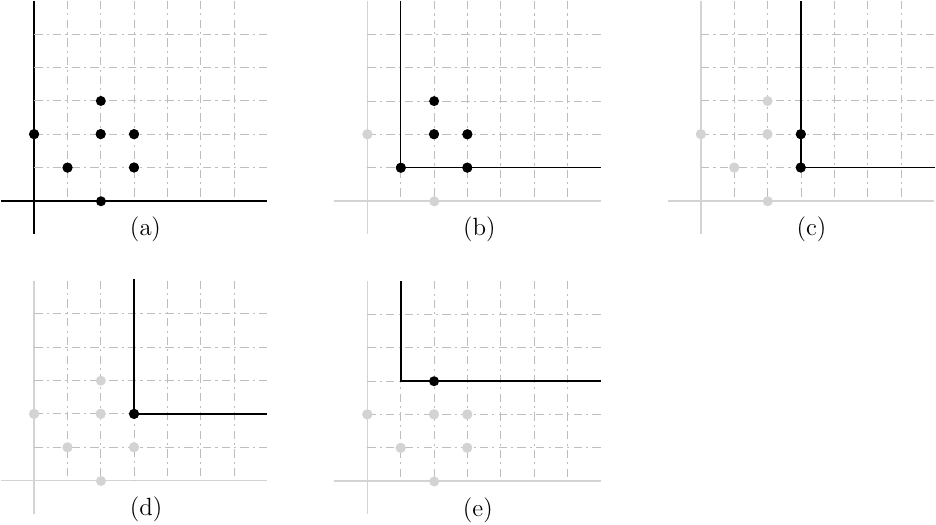}
	\caption{{The figure illustrates possible scenarios for PMI recurrence as used in the dynamic program with two leaders. Assume the origin of these figures to be $(0,0)$.
	In the case (a) there are separate points along both coordinates, so we have $A_{0,0} = \max (A_{0,1}+1, A_{1,0}+1 )$. In the case (b) there are two points along $x$ and one point along $y$, so we have $A_{1,1} = \max (A_{1,2}+1, A_{2,1}+1 )$. Note that point $(1,1)$ is minimum along both $x$ and $y$ coordinates.
	 In the case (c) there are two points along $y$ and one point along $x$, so  $A_{3,1} = \max (A_{3,2}+1, A_{4,1}+1 )$. In Figure (d), there is a point along $x$ and a point along $y$ (same point in this case), so we have $A_{3,2} = \max (A_{3,3}+1, A_{4,2}+1 )$. In case (e) there is a point in $x$ coordinate but no point along $y$, so $A_{1,3} = \max (A_{1,4}+1, A_{3,3}+0 )$.
	 }
	}

	\label{fig:different_A}

\end{figure}

In the absence of conflict, Observation~\ref{obs1} guarantees that we can start our sequence with any point with the unique minimum value in some fixed coordinate. However, as suggested by Lemma~\ref{obs:conflict_bound}, in case of a conflict, we cannot include all points to PMI. Thus, we need to include some of the points and exclude others. The longest PMI sequence can be found by computing $m$ subsequences corresponding to $m$ coordinates and taking the maximum. We conclude that $\alpha^{[c_1,c_2,\ldots,c_m]}$ can be obtained by the following recurrence:

\vspace{-0.13in}

\begin{equation}
\label{eq:alpha}
\alpha^{[c_1,c_2,\ldots,c_m]} = \max_{1\le j\le m}( \alpha^{[c_1,c_2,\ldots,c_j+1,\ldots,c_m]} + \mathbf{1}_{c_j}  ),
\end{equation}

where
\begin{equation}
\label{eq:1c}
\small
\mathbf{1}_{c_j} = \left\lbrace\begin{array}{cc}
    1 & \text{if}\;\exists\; D_p \text{ s.t. } [D_p]_j=c_j \text{ and } [D_{p}]_{j}'\ge c_{j'}, \forall {j'}\ne j. \\
    0 & \text{otherwise.}
\end{array}
\right.
\end{equation}
 We plan to pre-compute and memoize all {\em required} values of $\alpha^{[c_1,\ldots c_d]}$ in a table. Clearly there are
infinitely many possible values for $c_j$; however, we observe the following:

\begin{observation}
	\label{obs:empty_space}
	Let $\L^j_{i,j}$ and $\L^j_{i,j+1}$ be the $j^{th}$ coordinate values of two consecutive points in $\L^j$, then
	$$
	\alpha^{[c_1,c_2,\ldots,x,\ldots, c_m]} = \alpha^{[c_1,c_2,\ldots,\L^j_{i,j+1},\ldots, c_m]},
	$$
	for all $x$, such that $\L^j_{i,j} < x \le \L^j_{i,j+1}$.
\end{observation}

Observation~\ref{obs:empty_space} implies that there are at most $n$ different values for each variable $c_j$, which gives at most $n$ unique values for $\alpha^{[c_1,c_2,\ldots,c_m]}$. Thus, we only keep a table of size $n^m$ for computation and storage of solutions to all sub-problems. For some intuition on the working of dynamic program in Algorithm~\ref{algo:DP}, we refer the reader to Figure~\ref{fig:different_A}.

\begin{algorithm}[htb]
	\caption{PMI - Dynamic Program}
	\label{algo:DP}
	\begin{algorithmic}[1]

		\Procedure{\texttt{PMI-DP} }{$\L^1,\L^2,\ldots,\L^m$}

		\State $z_j$ be number of unique values  of $j^{th}$ coordinate among all points.
		\State $z = \max(z_1,z_2,\ldots,z_m)$
		\State Define a $m$-dimensional array $A$ with dimensions
		$(z+1)\times (z+1)\times\ldots (z+1) $
		\State Let $A_{c_1,c_2,\ldots, c_m}$, i.e. value of $A$ at index set ${c_1,c_2,\ldots, c_m}$ represents
		$\alpha^{[c_1,c_2,\ldots,c_m]}$ as in \eqref{eq:alpha}.
        \For{$k$ from $1$ to $m$}
		    \State $A_{c_1,c_2,\ldots, c_m} \gets 0$ for $c_k = z, c_{k'} \le z, k'\ne k$.
	    \EndFor
		\For{$j$ from $z-1$ to $0$}
        \For{$k$ from $1$ to $m$}
		\State Compute $A_{c_1,c_2,\ldots, c_m}$ for $c_k = j,c_{k'} \le j, k'\ne k$ using \eqref{eq:alpha}.
		\EndFor
		\EndFor
		\State \textbf{return} $A_{0,0,\ldots,0}$
		\EndProcedure

	\end{algorithmic}
\end{algorithm}

\textcolor{blue}{We now state and prove the main result of this section:
 \begin{theorem}
 	\label{thm:dynamic}
Given a graph $G$ on $n$ vertices, and $m$ leaders, Algorithm~\ref{algo:DP}  returns a longest PMI sequence of distance-to-leaders vectors in ${O(m(n\log n + n^m))}$ time.
 \end{theorem}
\textit{Proof:}
The correctness of Algorithm~\ref{algo:DP} follows from Observation~\ref{obs1}, Observation ~\ref{obs:empty_space}, and Lemma~\ref{obs:conflict_bound} so all that remains is to prove the time complexity. Computing sorted lists $\L^1,\L^2,\ldots,\L^m$ takes $O(m n\log n)$ time.
Each value of $A_{c_1,c_2,\ldots, c_m}$ can be computed by taking the maximum of $m$ known values previously computed, and saved in multi-dimensional array $A$. The value of $\mathbf{1}_{c_j}$  can be computed in constant time by checking whether element at the last index of $\L^j$ has $j^{th}$ coordinate equal to $c_j$ as defined in \eqref{eq:1c}. The  multi-dimensional array $A$ contains at most $n^{m}$ values at the completion each of which takes constant amount of time to compute. Therefore running time of this algorithm is bounded by $O(m (n\log n + n^m))$.
\qed
}

Appendix illustrates the algorithm through an example.

\begin{remark}
We note that an exact algorithm to compute the longest PMI sequence in $O(m^n)$ was proposed in \cite{yaziciouglu2016graph}. Since $m$ is much smaller than $n$ typically, the dynamic programming solution in Algorithm~\ref{algo:DP} computes the longest PMI sequence in a much lesser $O(m (n\log n + n^m))$ time.
\end{remark}

\section{Linearithmic Time Approximation Algorithm for the Distance-based Bound}
\label{sec:comp_greedy}
In this section, we discuss a greedy algorithm that takes linearithmic time to approximate the lower bound $\delta(G,V_\ell)$. 
The algorithm gives very close approximations in practice, as illustrated numerically in Section~\ref{sec:eval}. We also discuss the approximation guarantees of the algorithm. 

The main idea behind the greedy algorithm is to make {\em locally} optimal choices when faced with the situation in Lemma~\ref{obs:conflict_bound}, that is, when including a point in PMI results in discarding a subset of points from possible future consideration. In this case, the best thing to do locally is to pick a point that results in the loss of the minimum number of other points. The details are provided in Algorithm~\ref{algo:greedy}. We also provide an illustration of this algorithm in the Appendix.
\begin{algorithm}[htb]
	\caption{PMI-Greedy Algorithm}
	\label{algo:greedy}
	\begin{algorithmic}[1]

		\Procedure{\texttt{PMI-Greedy}}{$\L^1,\L^2,\ldots,\L^m$}

		\State $\calD \gets \emptyset$\Comment{Initially empty sequence}
		\While{$\L^1 \ne \emptyset$ }

		\State $X_j\gets \{\L^j_i: \L^j_{i,j}=L^j_{1,j} \}$ for all $j$.
		\If{$\exists j$ such that $|X_j|=1$}\Comment{Unique min.}
		\State $\calD \gets [\calD\;\;X_j]$
		\State Remove $X_j$ from all lists.
		\Else
		\State Let $j' \gets \argmin_j |X_j|$\Comment{Get smallest $X_j$}
		\State $\calD \gets [\calD\;\;\L^{j'}_1]$
		\State Remove all points in $X_{j'}$ from all lists.
		\EndIf
		\EndWhile
		\State \textbf{return}  $\calD$
		\EndProcedure

	\end{algorithmic}
\end{algorithm}

\begin{prop}
Algorithm~\ref{algo:greedy} computes a PMI sequence in $O(m n\log n)$ time. The length of the PMI sequence returned by the algorithm is a $\min(m,\frac{n}{m})$-approximation to the optimal length, where $m$ is the number of leaders and $n$ is the total number of nodes. Further, the approximation ratio of PMI lengths is $\log n$ if $m\le \log n$ or $m\ge \frac{n}{\log n}$.
\end{prop}
\textit{Proof:}
Regarding the time complexity, computing sorted lists $\L^1,\L^2,\ldots,\L^m$ takes $O(m\times n\log n)$ time.
Once we have $m$ sorted lists, we can keep the indices and count of points with the minimum coordinate value in ($m+1$) Min-Priority queues ($m$ queues to maintain lists $\L^1,\L^2,\ldots,\L^m$ and one queue for $X_1,X_2,\ldots,X_m$).
Cost of one \textit{update}, or \textit{delete} operation is $O(\log n)$ in a Priority queue. Since we will \textit{update} and/or \textit{delete} at most $n$ points from $m$ queues. In total we will perform at most $m\times n$ deletions and $m\times n$ updates, thus, the overall time complexity is $O(m n\log n)$.

Regarding the $m$-approximation ratio, we observe that there are at least $\frac{n}{m}$ different values in at least one coordinate. Otherwise, we may assume that we have at most $\frac{n}{m} -1$ unique values in each coordinate. This would imply that there are at most $(\frac{n}{m} -1)\times m$ distinct points by the pigeonhole principle, which contradicts that we have $n$ unique points. As Algorithm~\ref{algo:greedy} picks all distinct values in any coordinate, the returned PMI sequence has a size of at least $\frac{n}{m}$. Note that there is at least one unique minimum point (corresponding to the leader itself) in each of $m$ directions, so the algorithm is $(\frac{n}{m})$-approximation as well.

It is evident that when $m$ is at most $\log n$, there are at least $\frac{n}{\log n}$ different values in at least one coordinate by the same argument. To see why $\log n$-approximation ratio holds when $m$ is large, note that there is at least one unique minimum point (corresponding to the leader itself) in each of $m$ directions so when $m\ge \frac{n}{\log n}$, the algorithm will include all of those unique points in the returned PMI sequence.
 Thus, the approximation ratio follows.
\qed

\begin{figure}[htb]
\centering
\includegraphics[scale=0.53]{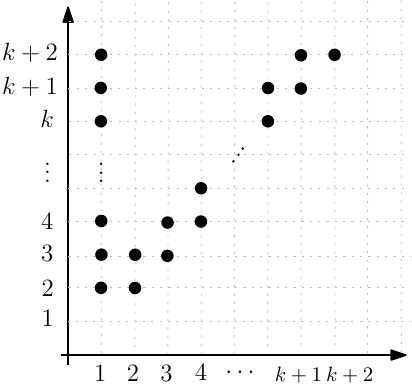}
\caption{Example discussed in Remark \ref{remark:greedy}.}
\label{fig:bad_example}
\end{figure}

If there exists a PMI sequence of length $n$, then the network is strong structurally controllable with a given set of leaders. The greedy algorithm presented above always returns a PMI sequence of length $n$ if there exists one. 
\begin{lem}
If there exists a PMI sequence of length $n$, then Algorithm~\ref{algo:greedy} always returns an optimal PMI.
\end{lem}
\textit{Proof:}
We observe that if there exists a PMI sequence of length $n$, then by Lemma~\ref{obs:conflict_bound}, we
can not have a conflict as defined in Section~\ref{sec:exact_algorithm}. In the absence of any conflict, we can always find a unique minimum point along some coordinate. Consequently, in Algorithm~\ref{algo:greedy}, statements in \textbf{else} will never be executed and algorithm will return a PMI sequence of length $n$.
\qed
\begin{remark}
\label{remark:greedy}
While in many cases, Algorithm~\ref{algo:greedy} achieves a solution close to optimum, we observe that examples can be constructed for which a greedy solution may not be globally optimal. In the example outlined in Figure~\ref{fig:bad_example}, we have two leaders and points are placed at
$S = \{ (2,2),(2,3), (3,3), (3,4),(4,4),\ldots, (k+1,k+1),(k+1,k+2),(k+2,k+2)\}$ and at $T = \{ (1,2),(1,3),(1,4),\ldots, (1,k+2)\}$. An optimal PMI has all $2k$ points while Algorithm~\ref{algo:greedy} above may only pick $k+3$ points.
\end{remark}

\section{Application: Leader Selection for Strong Structural Controllability}
\label{sec:Application}
In this section, we briefly discuss an application of computing the distance-based bound in approximately solving a leader selection problem for strong structural controllability, which is intractable to solve exactly \cite{chapman2013strong, Bhangale2015complexity,van2017distance}. The problem of finding the minimum number of leaders to make a network strong structurally controllable is known to be NP-complete \cite{chapman2013strong,van2017distance}. Here, we consider the problem of finding a set $V_\ell$ of $m$ leaders that maximizes the dimension of SSCS, i.e., 
 \begin{equation}
 \label{lselect}
 \begin{aligned}
 & \underset{V_\ell \subseteq V}{\text{maximize}}
 & & \gamma(G,V_\ell);
 & \text{subject to}
 & & |V_\ell| = m.  \\
 \end{aligned}
\end{equation}

In light of \eqref{eq:PMI_bound}, the distance-based bound $\delta(G,V_\ell)$ can be used to obtain an approximate solution to such a leader selection problem by solving
 \begin{equation}
 \label{lselect2}
 \begin{aligned}
 & \underset{V_\ell \subseteq V}{\text{maximize}}
 & & \delta(G,V_\ell);
 & \text{subject to}
 & & |V_\ell| = m.  \\
 \end{aligned}
\end{equation}

Any solution to the problem in \eqref{lselect2}, $V_\ell^*$, ensures that the resulting dimension of SSCS is at least $\delta(G,V_\ell^*)$. While the problem in \eqref{lselect2} is still hard to solve due to its combinatorial nature, an approximate solution can be obtained by utilizing an algorithm for computing $\delta(G,V_\ell)$.  We present a simple greedy heuristic for leader selection for strong structural controllablity using PMI sequences of distance-to-leaders vectors in a graph. Given a network $G$ and the number of leaders $m$ as inputs, the main idea is to iteratively select leaders that maximally increase the length of resulting PMI sequences. The outline of the heuristic is in Algorithm \ref{algo:leader_select}.

\begin{algorithm}[!h]
	\caption{Greedy Leader Selection Algorithm}
	\label{algo:leader_select}
	\begin{algorithmic}[1]

		\Procedure{\texttt{Leader-Select}}{$G,m$}
		\State $V_\ell \gets \emptyset$, $V'\gets V$
        \For{$i \gets 1$ to $m$}
        \For {each $v\in V'$}
        \State Compute PMI sequence with $V_\ell \cup \{v\}$ leaders.
        \EndFor
        \State Choose $v'\in\ V'$ that gives a PMI sequence of maximum length with $V_\ell \cup \{v'\}$ leaders.
        \State $V_\ell \gets V_\ell \cup \{v'\}$.
        \State $V'\gets V'\setminus \{v'\}$.
        \EndFor
		\EndProcedure

	\end{algorithmic}
\end{algorithm}


The time complexity of Algorithm \ref{algo:leader_select} depends on the complexity of computing a PMI sequence with a given set of leaders. Using the algorithm in \cite{yaziciouglu2016graph} to compute PMI sequences, the complexity of Algorithm \ref{algo:leader_select} is $O(n\times m^n)$, which means the algorithm is practically infeasible even for $m=2$. Using Algorithm \ref{algo:DP} (dynamic programming) to compute PMI sequences, \textcolor{blue}{the time complexity of Algorithm~\ref{algo:leader_select} is $O(n^2\log n + m\times n^{m+1})$, which is a significant improvement. The first term in this expression is the cost of sorting the distance lists and the second term is the cost of computing $n$ PMI sequences when the leader set includes $m$ nodes (the iteration when $i=m$) as this dominates the cost of all previous iterations.} However, leader selection in Algorithm \ref{algo:leader_select} can be achieved in $O(m^2n^2\log n)$ time if we use Algorithm \ref{algo:greedy} to compute PMI sequences. Section~\ref{sec:numerical_leader_selecet} provides numerical evaluation of the leader selection algorithm (Algorithm \ref{algo:leader_select}) that uses exact/greedy algorithm to compute $\delta(G,V_\ell)$ to approximately solve \eqref{lselect2}. 
\section{Bounds in Paths and Cycles}
\label{sec:PathCycle}
In this section, we explore connections between graph-theoretic properties and the length of the longest PMI in path and cycle graphs. As a result, we show interesting topological bounds on the dimension of SSCS in such graphs with a given set of leader nodes. We note that our results differ from previous works in this direction in two aspects: first, we specifically study the strong structural controllability of such graphs; second, instead of focusing on complete controllability, we provide tight bounds on the dimension of SSCS even when the graph is not strong structurally controllable with a given set of leaders (e.g., \cite{parlangeli2012reachability,liu2018controllability}).

Recall that a node with a single neighbor is called \textit{leaf}. Moreover, given $G=(V,E)$ and $V'\subset V$, then the subgraph of $G$ \emph{induced} on $V'$ is the graph whose vertex set is $V'$, and the edge set consists of all of the edges in $E$ that have both endpoints in $V'$. We start with the following obvious fact.
\begin{fact}
A path graph in which a leaf is a leader has a PMI sequence of length $n$.
\label{fact:leaf_leader}
\end{fact}
\begin{theorem}
Let $G$ be a path graph on $n$ nodes, let $V_\ell$ be a set of $m\le n$ leader nodes, and let $G^{-{V_\ell}}$ denote the subgraph of $G$ induced on vertices $V\setminus V_\ell$. Then the following holds:
\begin{enumerate}
    \item[(i)] If the number of connected components in $G^{-{V_\ell}}$ is less than $m+1$, then the longest PMI sequence induced by ${{V_\ell}}$ has length $n$.
    \item[(ii)] If the number of connected components in $G^{-{V_\ell}}$ is $m+1$, then ${{V_\ell}}$ induces a PMI sequence of length $n-a$, where $a$ is the size of the smallest connected component in $G^{-{V_\ell}}$.
\end{enumerate}
\label{claim:path_full_pmi}
\end{theorem}
\textit{Proof:}
(i) Removal of a node from a path results in at most two connected components. Hence, $G^{-{V_\ell}}$ has at most $m+1$ such components. If the number of components is less than $m+1$, either one of the leader nodes is a leaf, or at least two leaders are adjacent. If a leaf $x$ is chosen as a leader, then by Fact~\ref{fact:leaf_leader}, we can get a PMI sequence of length $n$. Assuming none of the leaders is a leaf node, let $v_i$ and $v_{i+1}$ be adjacent leader nodes; further assume that $i<n/2$ without loss of generality. We will construct a PMI sequence of length $n$ based on these two leaders as follows.
\begin{multline*}
\footnotesize
\left[ \begin{array}{cccccccc}
\left[ \begin{array}{c}  0 \\ 1 \end{array}\right] ,
\left[ \begin{array}{c}  1 \\ 0 \end{array}\right] ,
\left[ \begin{array}{c}  1 \\ 2 \end{array}\right] ,
\left[ \begin{array}{c}  2 \\ 1 \end{array}\right] ,
\cdots,
\left[ \begin{array}{c}  i-1  \\ i \end{array}\right] ,
\left[ \begin{array} {c} i  \\ i-1 \end{array}\right] ,\\
\left[ \begin{array} {c} i+1  \\ i \end{array}\right] ,
\left[ \begin{array} {c} i+2  \\ i+1 \end{array}\right] ,
\cdots,
\left[ \begin{array}{c}  n-i  \\ n-i-1 \end{array}\right]
\end{array}\right]
\end{multline*}

(ii) If the smallest connected component $X$ contains either of the leaf nodes, then $G^{-X}$ has a leaf leader node and thus has a PMI sequence of length $n-|X|$ by Fact~\ref{fact:leaf_leader}. If $X$ doesn't contain leaf nodes, then there exist two leader nodes $v_i,v_j$ are adjacent to some nodes in $X$.
Also, assume that $v_i$ is not farther away from a leaf node than $v_j$ is.
Then, the following sequence of distance-to-leaders vectors defines a PMI sequence of claimed length.
\begin{multline*}
\footnotesize
\left[ \begin{array}{cccccccc}
\left[ \begin{array}{c}  0 \\ a+1 \end{array}\right] ,
\left[ \begin{array}{c}  a+1 \\ 0 \end{array}\right] ,
\left[ \begin{array}{c}  1 \\ a+2 \end{array}\right] ,
\left[ \begin{array}{c}  a+2 \\ 1 \end{array}\right] ,
\cdots, \\
\left[ \begin{array}{c}  i-1  \\ a+i \end{array}\right] ,
\left[ \begin{array} {c} a+i  \\ i-1 \end{array}\right] ,
\left[ \begin{array} {c} a+i+1  \\ i \end{array}\right] ,
\left[ \begin{array} {c} a+i+2  \\ i+1 \end{array}\right] ,
\cdots,\\
\left[ \begin{array}{c}  n-i  \\ n-i-a-1 \end{array}\right]
\end{array}\right]
\end{multline*}

\qed

\begin{theorem}
Let $G$ be a cycle on $n$ nodes, let $V_\ell$ be a set of $2\le m\le n$ leader nodes, and let $G^{-{V_\ell}}$ denote the subgraph of $G$ induced on vertices $V\setminus V_\ell$. Then, the following holds:
\begin{enumerate}
    \item[(i)] If the number of connected components in $G^{-V_\ell}$ is less than $m$, then the longest PMI sequence induced by $V_\ell$ has length $n$.
    \item[(ii)] If the number of connected components in $G^{-{V_\ell}}$ is exactly $m$, then ${{V_\ell}}$ induces a PMI sequence of length $n-a$, where $a$ is the size of the smallest connected component in $G^{-{V_\ell}}$.
\end{enumerate}

\label{claim:cycle_full_pmi}
\end{theorem}
\textit{Proof:}
(i) Removing a single node from a cycle does not affect the number of connected components. However, the removal of every subsequent node will result in at most one extra component. Thus, the total number of connected components is at most $m$ after the removal of $m$ nodes. If the number of components is less than $m$ in $G^{-{V_\ell}}$, then at least two nodes in ${V_\ell}$ are neighbors in $G$. Let $v_1$ and $v_{2}$ be an arbitrary adjacent pair in ${V_\ell}$. We will construct a PMI sequence of length $n$ based on these two leaders. Consider the nodes in $G$ with the following distance-to-leaders vectors:
\begin{equation*}
\footnotesize
\left[ \begin{array}{cccccccc}
\left[ \begin{array}{c}  0 \\ 1 \end{array}\right] ,
\left[ \begin{array}{c}  1 \\ 0 \end{array}\right] ,
\left[ \begin{array}{c}  1 \\ 2 \end{array}\right] ,
\left[ \begin{array}{c}  2 \\ 1 \end{array}\right] ,
\cdots,
\left[ \begin{array}{c}  \frac{n}{2}-1  \\ \frac{n}{2} \end{array}\right] ,
\left[ \begin{array} {c} \frac{n}{2}  \\ \frac{n}{2}-1 \end{array}\right]
\end{array}\right]
\end{equation*}
when $n$ is even, and
\begin{equation*}
\footnotesize
\left[ \begin{array}{cccccccc}
\left[ \begin{array}{c}  0 \\ 1 \end{array}\right] ,
\left[ \begin{array}{c}  1 \\ 0 \end{array}\right] ,
\left[ \begin{array}{c}  1 \\ 2 \end{array}\right] ,
\left[ \begin{array}{c}  2 \\ 1 \end{array}\right] ,
\cdots,
\left[ \begin{array} {c} \lfloor{\frac{n}{2}}\rfloor  \\ \lfloor{\frac{n}{2}}\rfloor \end{array}\right]
\end{array}\right]
\end{equation*}
when $n$ is odd. This defines a PMI sequence of length $n$.

(ii) An argument identical to proof of Theorem~\ref{claim:path_full_pmi}(ii) can be used here to prove (ii) as well.
\qed
Theorems~\ref{claim:path_full_pmi} and \ref{claim:cycle_full_pmi} imply graph-theoretic bounds on the dimension of SSCS for path and cycle graphs. A path (cycle) graph is strong structurally controllable with $V_\ell$ leaders if $G^{-{V_\ell}}$ has $m+1$ components ($m$ components in a cycle). Another direct implication of the above results is as follows.
\begin{cor}
Let $G$ be a path or cycle graph and let $V_\ell$ be a set of leaders, then the dimension of SSCS is at least $n-a$, where $a$ is the smallest distance between any two leader nodes.
\end{cor}

\section{Numerical Evaluation}
\label{sec:eval}
In this section, we numerically evaluate our results on Erd\"os-R\'enyi (ER) and Barab\'asi-Albert (BA) graphs. In ER graphs, any two nodes are adjacent with a probability $p$. BA graphs are obtained by adding nodes to an existing graph one at a time. Each new node is connected to $\varepsilon$ existing nodes with probabilities proportional to the
degrees of those nodes.
\subsection{Comparison of Algorithms}
\label{sec:DP_GReedy_Eval}
\begin{figure*}[!h]
\centering
\begin{subfigure}{0.225\linewidth}
\centering
\includegraphics[scale=0.3]{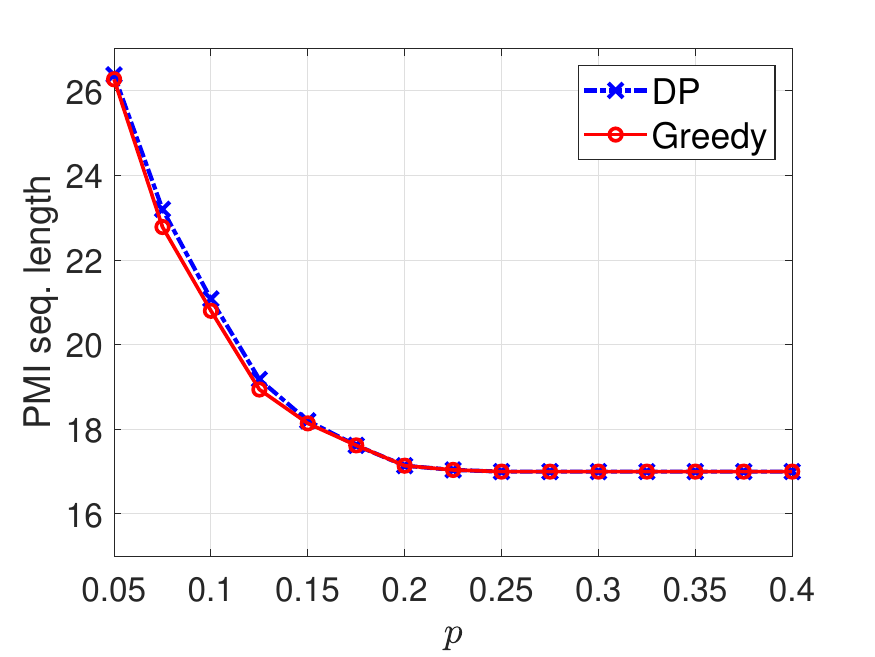}
\caption{ER}
\end{subfigure}
\begin{subfigure}{0.225\linewidth}
\centering
\includegraphics[scale=0.3]{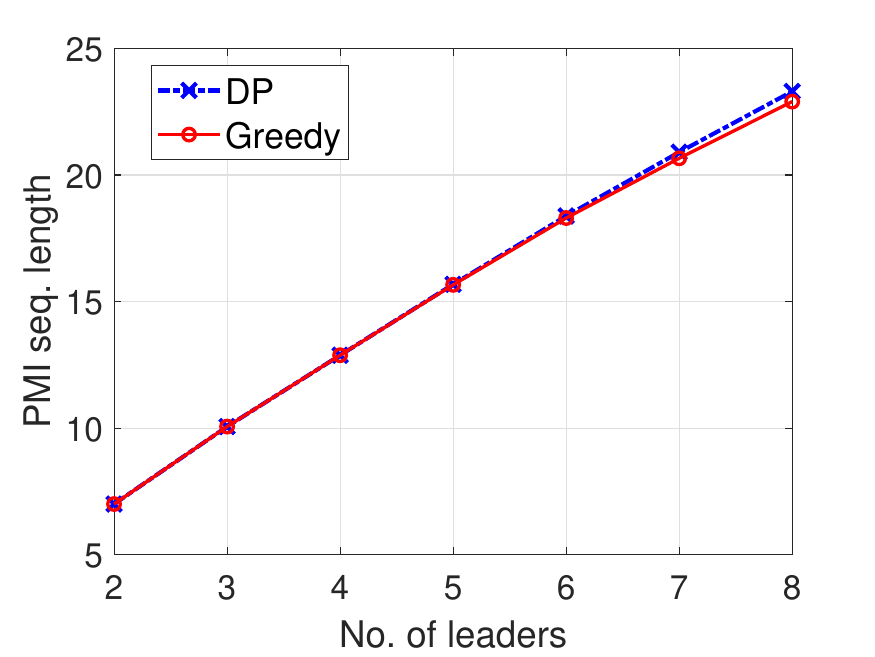}
\caption{ER}
\end{subfigure}
\begin{subfigure}{0.225\linewidth}
\centering
\includegraphics[scale=0.3]{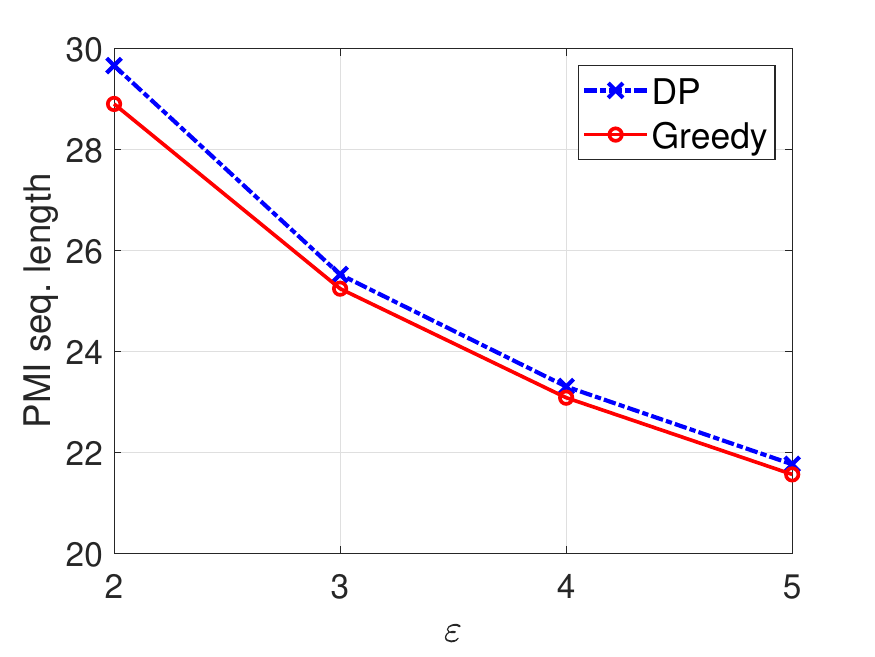}
\caption{BA}
\end{subfigure}
\begin{subfigure}{0.225\linewidth}
\centering
\includegraphics[scale=0.3]{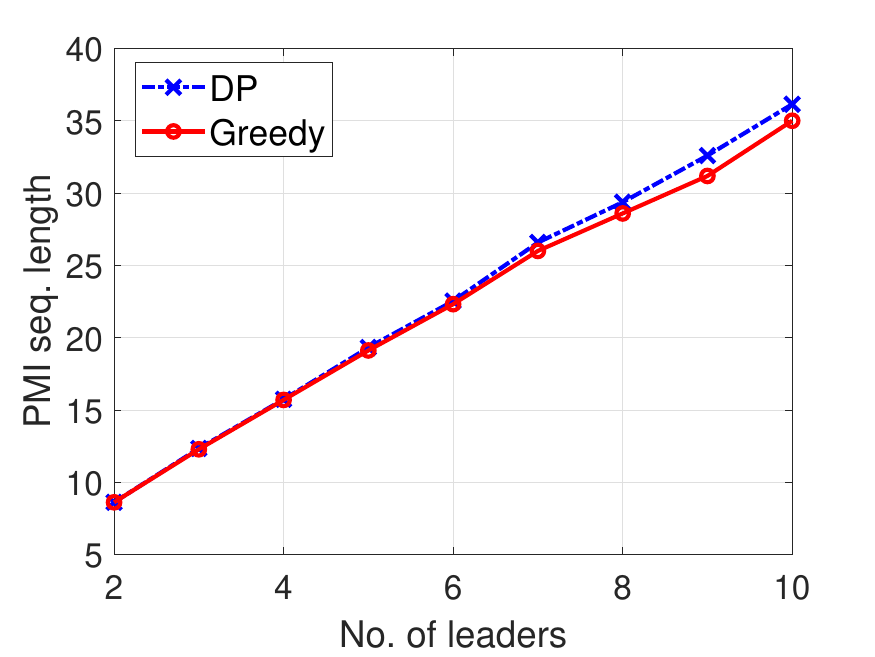}
\caption{BA}
\end{subfigure}
\caption{Comparison of Algorithm~\ref{algo:DP} (dynamic programming) and Algorithm~\ref{algo:greedy} (greedy) for computing the distance-based bound.}
\label{fig:DP_Greedy}
\end{figure*}

First, we compare the performance of the exact dynamic programming algorithm (Algorithm~\ref{algo:DP}) and the approximate greedy algorithm (Algorithm~\ref{algo:greedy}) for computing the maximum-length PMI sequences. For simulations, we consider graphs with $n=200$ nodes. For ER graphs, we first plot the length of PMI sequences computed by using Algorithms~\ref{algo:DP} and \ref{algo:greedy} as a function of $p$ while fixing the number of leaders (selected randomly) to be eight (Figure~\ref{fig:DP_Greedy}(a)). Second, we fix $p=0.075$, and plot the length of PMI sequences as a function of the number of leaders selected randomly (Figure~\ref{fig:DP_Greedy}(b)). We repeat similar plots for BA graphs in Figures~\ref{fig:DP_Greedy}(c) and \ref{fig:DP_Greedy}(d). We fix the number of leaders to be eight in Figure \ref{fig:DP_Greedy}(c) and set $\varepsilon=2$ in Figure~\ref{fig:DP_Greedy}(d). Each point in the plots in Figure~\ref{fig:DP_Greedy} corresponds to the average of 50 randomly generated instances. From the plots, it is clear that the greedy algorithm, which is much faster than the DP algorithm, performs almost as good as the DP algorithm. The length of PMI sequences returned by the greedy algorithm is very close to the length of the longest PMI sequences.

\subsection{Comparison of Bounds}
\label{sec:PMI_ZFS_Eval}
Next, we numerically compare our distance-based bound with another well known bound on the dimension of SSCS based on the notion of zero-forcing sets (ZFS) \cite{monshizadeh2014zero,monshizadeh2015strong}. First, we explain the notion of ZFS. Given a graph $G=(V,E)$ in which each node is colored either \textit{white} or \textit{black}, we repeatedly apply the following coloring rule: {\em If $v\in V$ is colored black and has one white neighbor $u$, then the color of $u$ is changed to black.} Now, given an initial set of black nodes (called \textit{input set}) in $G$, \emph{derived set} $V'\subseteq V$ is the set of all black nodes obtained after repeated application of the coloring rule until no color changes are possible. It is easy to see that for a given input set, the resulting derived set is unique. The input set is called a ZFS if the corresponding derived set contains all nodes in $V$. It is shown in \cite{monshizadeh2014zero,monshizadeh2015strong} that for a given set of leader nodes as input set, the size of the corresponding derived set is a lower bound on the dimension of SSCS.

\begin{figure}[!h]
\centering
\begin{subfigure}{0.53\linewidth}
\centering
\includegraphics[scale=0.2]{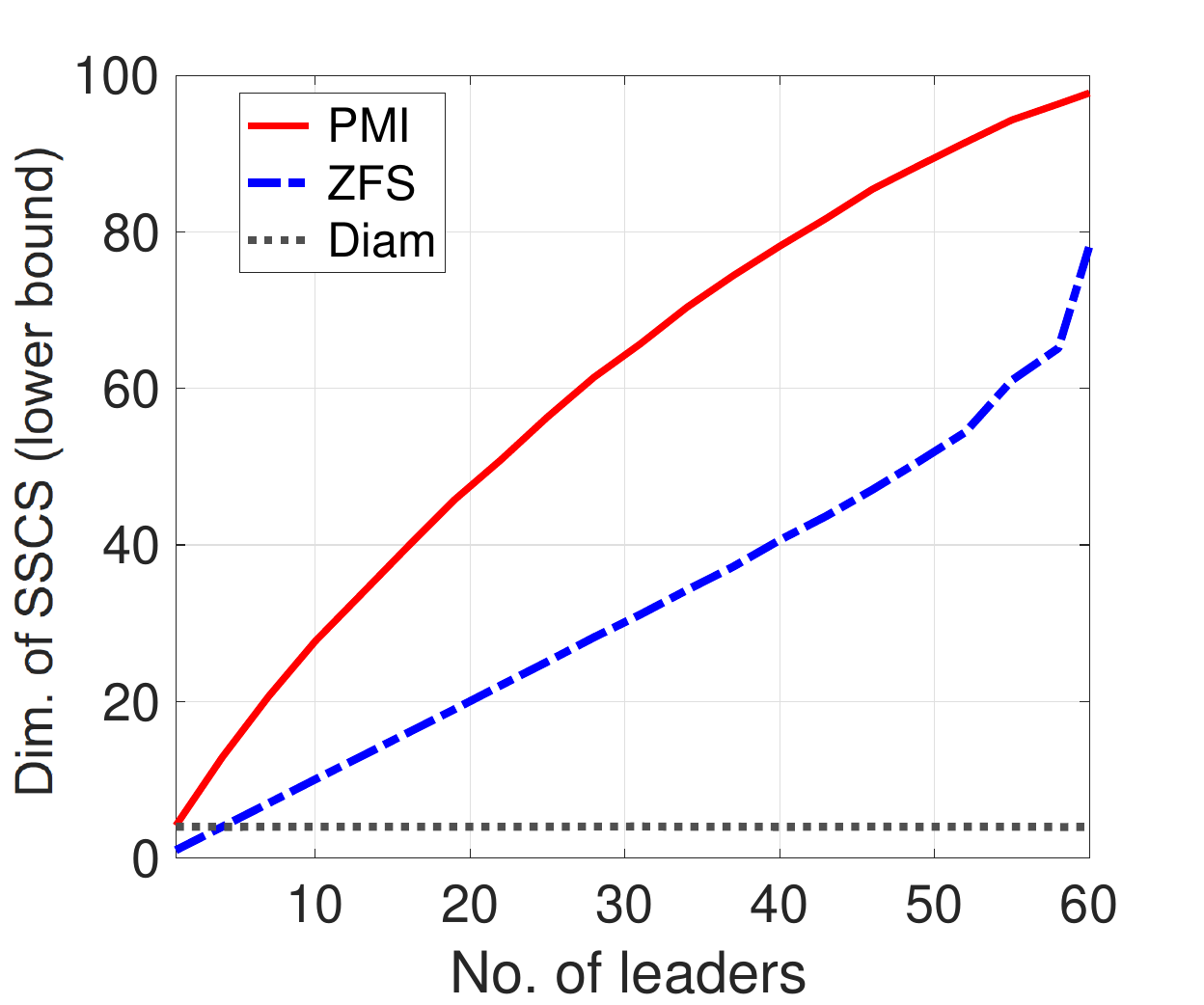}
\caption{ER ($p=0.1$)}
\end{subfigure}
\begin{subfigure}{0.43\linewidth}
\centering
\includegraphics[scale=0.2]{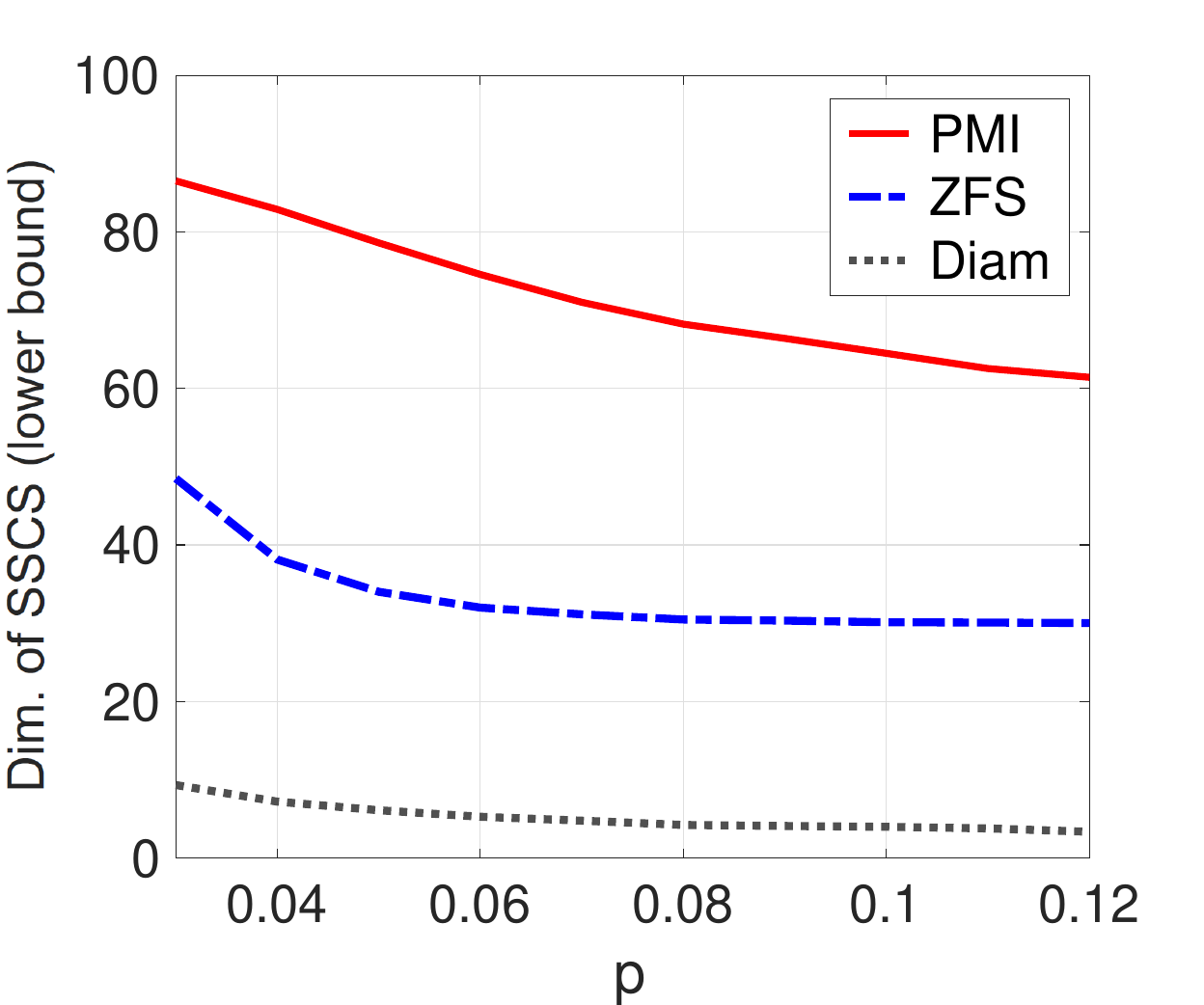}
\caption{ER ($|V_\ell|=30$)}
\end{subfigure}
\begin{subfigure}{0.53\linewidth}
\centering
\includegraphics[scale=0.2]{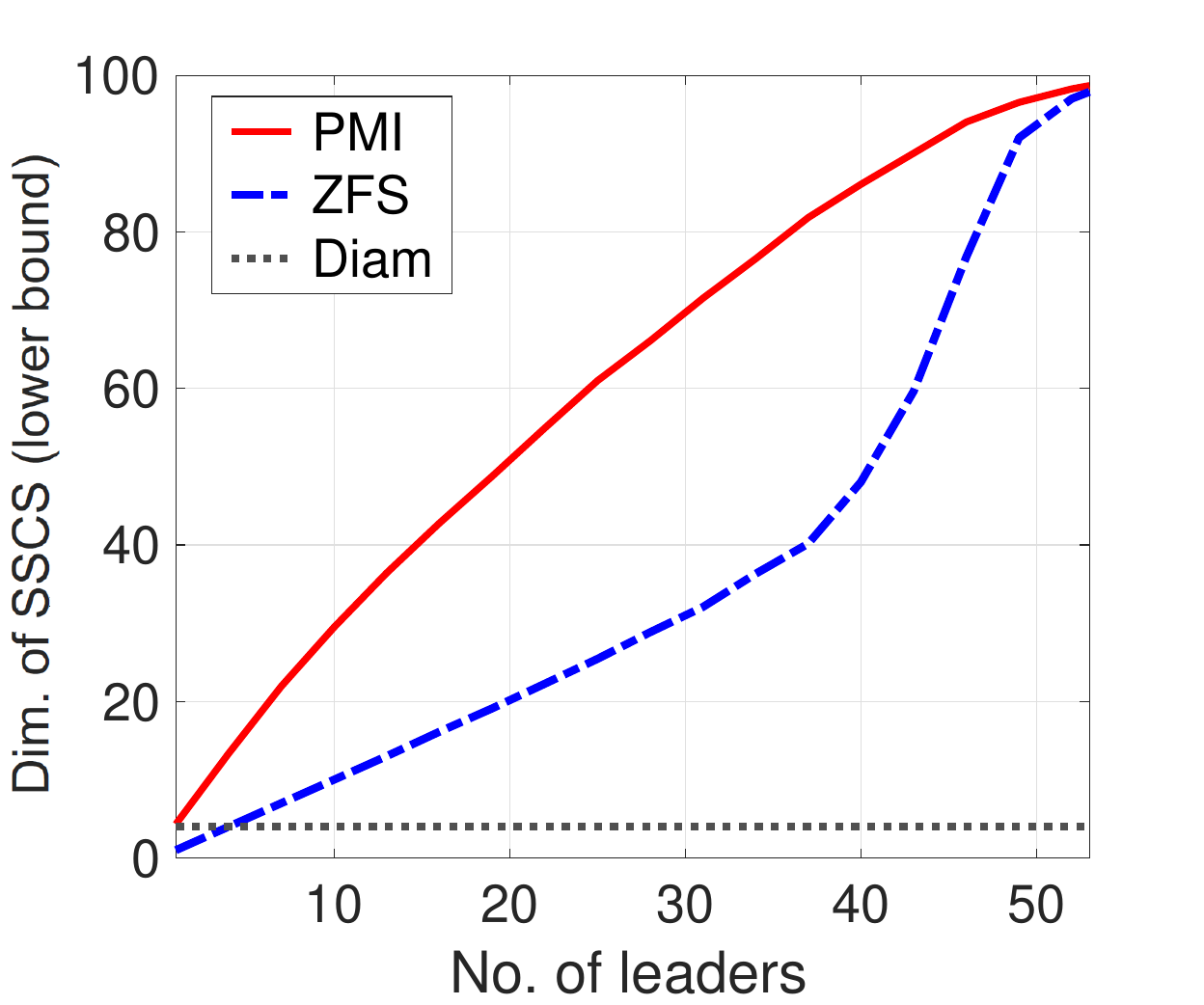}
\caption{BA ($\varepsilon=4)$}
\end{subfigure}
\begin{subfigure}{0.43\linewidth}
\centering
\includegraphics[scale=0.2]{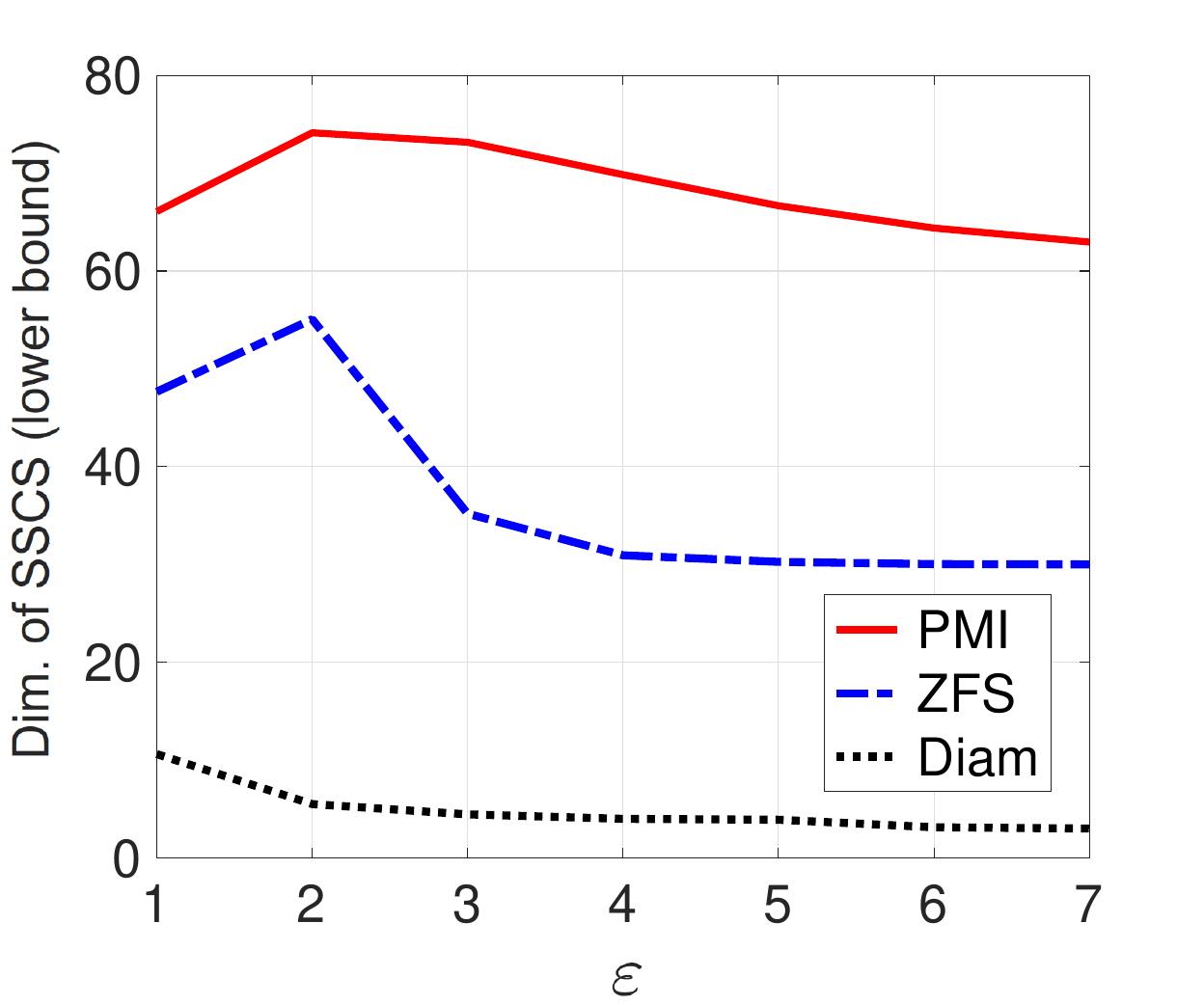}
\caption{BA ($|V_\ell|=30$)}
\end{subfigure}
\caption{Comparison of the distance-based and ZFS-based lower bounds on the dimension of SSCS in ER and BA graphs. The total number of nodes in all graphs is 100. The diameters of graphs (denoted by Diam) are also plotted.}
\label{fig:ERBA_Eval}
\end{figure}

In our simulations in Figure \ref{fig:ERBA_Eval}, for both ER and BA models, we consider graphs with $n=100$ nodes. 
In Figure~\ref{fig:ERBA_Eval}(a), we plot these bounds for ER graphs as a function of the number of leaders, which are selected randomly, while fixing $p=0.1$. 
Next, we fix the number of leaders to be 30 in Figure~\ref{fig:ERBA_Eval}(b) and plot bounds as a function of $p$.
As previously, each point in the plots is an average of 50 randomly generated instances. It is obvious that distance-based bound significantly outperforms the ZFS-based bound in all the cases.  
Similar results are obtained in the case of BA graphs, where 
we fix $\varepsilon=4$ in Figure~\ref{fig:ERBA_Eval}(c), and select the number of leaders to be 30 in Figure~\ref{fig:ERBA_Eval}(d). In all the plots, for a given set of leaders, lengths of PMI sequences are always greater than the derived sets; thus, distance-based bound on the dimension of SSCS is better than the one based on the derived sets.
\color{black}
\subsection{Leader Selection}
\label{sec:numerical_leader_selecet}
\begin{figure}[h!]
\centering
\begin{subfigure}{0.45\linewidth}
\centering
\includegraphics[scale=0.3]{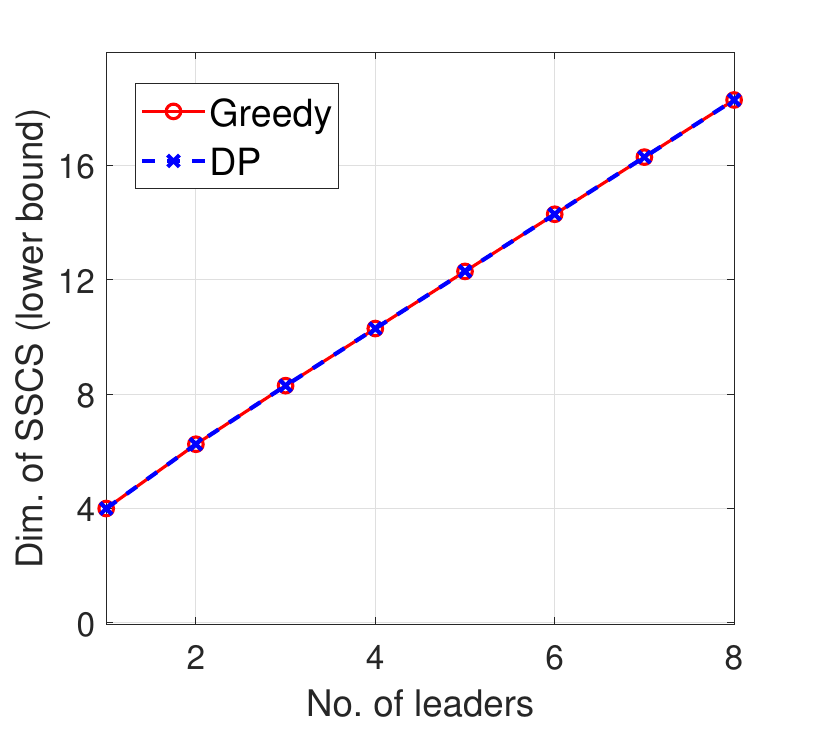}
\caption{ER ($p=0.3$)}
\end{subfigure}
\begin{subfigure}{0.45\linewidth}
\centering
\includegraphics[scale=0.3]{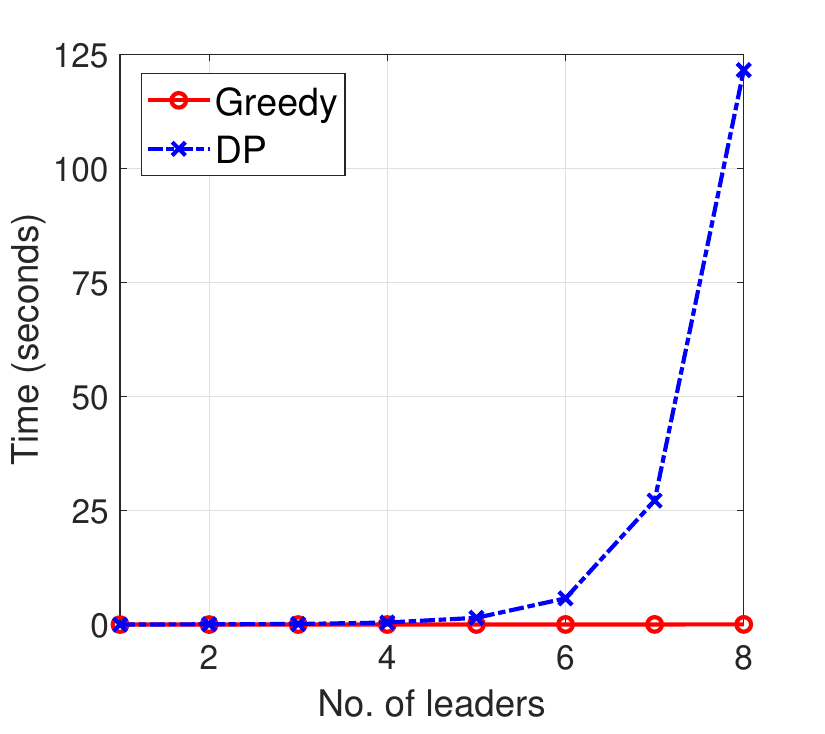}
\caption{ER ($p=0.3$)}
\end{subfigure}
\begin{subfigure}{0.45\linewidth}
\centering
\includegraphics[scale=0.3]{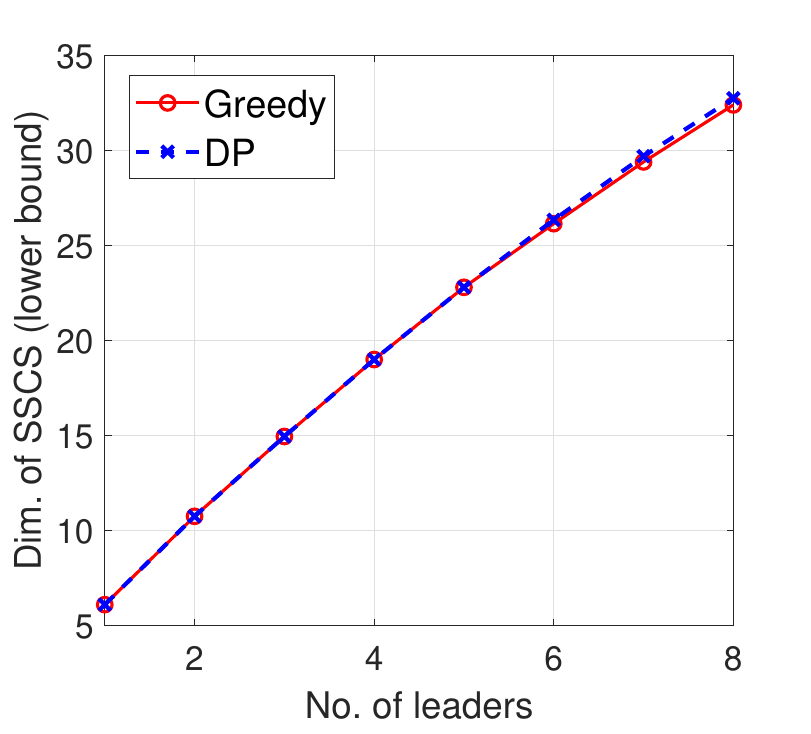}
\caption{BA ($\varepsilon=2$)}
\end{subfigure}
\begin{subfigure}{0.45\linewidth}
\centering
\includegraphics[scale=0.3]{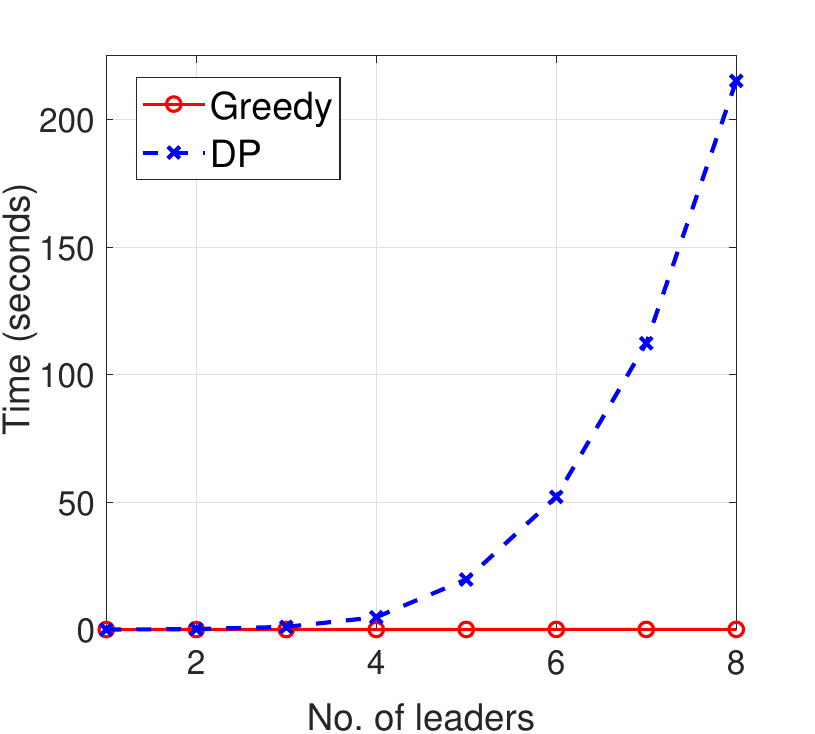}
\caption{BA $(\varepsilon=2$)}
\end{subfigure}
\caption{Comparison of the leader selection algorithm with the greedy computation and the exact computation (dynamic programming (DP)) of the distance-based bound. The total number of nodes in all graphs is 60.}
\label{fig:leader_select}
\end{figure}

We implement Algorithm~\ref{algo:leader_select} to illustrate the application of proposed algorithms to the leader selection problem given in~\eqref{lselect2}. Again, the networks were generated for both ER and BA models with $n = 60$ nodes. In order to compute the length of the longest PMI sequence, we use the bounds returned by the dynamic programming solution and the greedy algorithm. We compare the respective bounds on the dimension of SSCS and the computation times of the two algorithms. The results of our experiments are shown in Figure~\ref{fig:leader_select}. The resulting bounds are plotted against the number of leaders in Figures~\ref{fig:leader_select}(a) and \ref{fig:leader_select}(c), and the running times are plotted in Figures~\ref{fig:leader_select}(b) and \ref{fig:leader_select}(d). Each point in the plots corresponds to the average of $20$ runs. For both graph families, the bounds computed by the two algorithms are almost identical, but the greedy algorithm has the advantage of superior runtime that becomes more pronounced as the number of leaders increases.

\section{Conclusion}
\label{sec:conclusion}
We studied the computational aspects of a lower bound on the dimension of SSCS in networks with Laplacian dynamics. The bound is based on a sequence of distance-to-leaders vectors and has several applications. 
We proposed an algorithm that runs in $O(n^m)$ time (compared to $O(m^n)$ runtime of the algorithm in \cite{yaziciouglu2016graph}) to compute the bound.
We also presented a linearithmic approximation algorithm to compute the bound, which provided near-optimal solutions in practice. Further, we explored connections between the graph-theoretic properties and the distance-based bound in path and cycle graphs using the results. We plan to use these results to explore further the trade-offs between controllability and other desirable network properties, such as robustness and resilience to perturbations. We also believe that finding the longest PMI sequence of a given set of vectors is an interesting problem in its own respect as it naturally generalizes Erd\"os-Szekeres type sequences to higher dimensions \cite{erdos1935combinatorial,samotij2015number,linial2018monotone}.


\begin{figure*}[ht]
\centering
\includegraphics[scale=0.85]{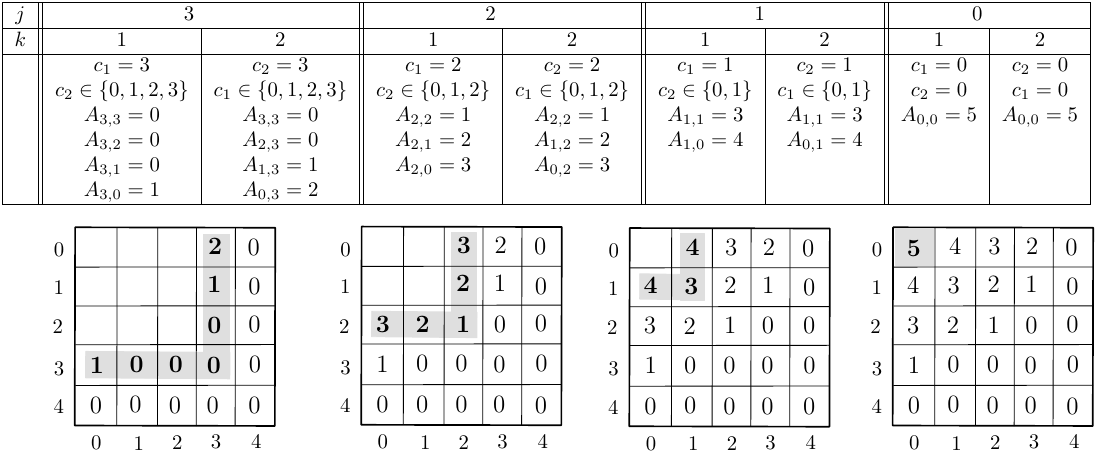}
\caption{Illustration of the run of Algorithm~\ref{algo:DP} for the graph in Figure~\ref{fig:Lap} with leaders $V_\ell = \{v_1,v_6\}$.}
\label{fig:Illustration_DP}
\end{figure*}

\appendix

\subsection*{Illustration of Dynamic Programming Algorithm (Algorithm~\ref{algo:DP})}
\label{sec:Appendix_DP}

For illustration, we consider the same graph as in Figure~\ref{fig:Lap} of the paper, where $V_{\ell} = \{v_1,v_6\}$. The example run is described in Figure~\ref{fig:Illustration_DP}. The values of $j$ and $k$ denote the loop variables in lines 9 and 10 of the Algorithm~\ref{algo:DP} respectively. In each iteration of $j$, one column and one row of the memoization variable $A$ is updated as shown in matrices in Figure~\ref{fig:Illustration_DP}. In fact, the value of each cell in the matrix is computed from values in the neighboring cells to the right and below using \eqref{eq:alpha}. For instance, $A_{1,0}$ is computed from $A_{1,1}$ (neighboring cell on the right) and $A_{2,0}$ (neighboring cell below). The first entry of the matrix, that is, $A_{0,0}$, returns the length of the longest PMI sequence.

\subsection*{Illustration of Greedy Algorithm (Algorithm~\ref{algo:greedy})}
\label{sec:Appendix_Greedy}

We illustrate it on the same graph as in Figure~\ref{fig:Lap} of the paper. Note that $\L^j$ is a list of points (distance-to-leaders vectors) that are sorted in a non-decreasing order with respect to the $j^{th}$ coordinate. In our example, such lists are given below.

\begin{multline*}
\small
\label{eq:L1}
\L^1 = \left[
\left[\begin{array}{c}0\\3\end{array}\right] ,\left[\begin{array}{c}1\\2\end{array}\right] ,\left[\begin{array}{c}1\\3\end{array}\right] ,\left[\begin{array}{c}2\\1\end{array}\right],\left[\begin{array}{c}2\\2\end{array}\right],\left[\begin{array}{c}3\\0\end{array}\right] \right];
\\
\L^2 = \left[
\left[\begin{array}{c}3\\0\end{array}\right] ,\left[\begin{array}{c}2\\1\end{array}\right] ,\left[\begin{array}{c}1\\2\end{array}\right] ,\left[\begin{array}{c}2\\2\end{array}\right],\left[\begin{array}{c}0\\3\end{array}\right],\left[\begin{array}{c}1\\3\end{array}\right]
\right].
\end{multline*}

\begin{figure}[h]
\centering
\includegraphics[scale=0.95]{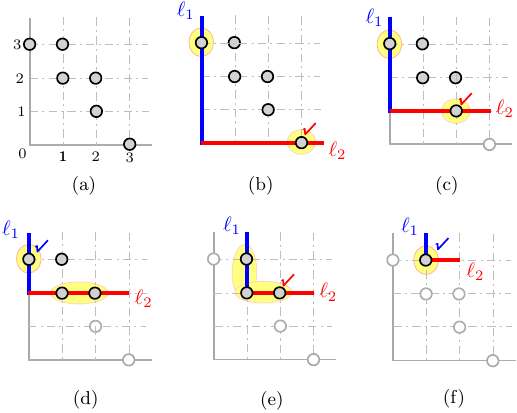}
\caption{Illustration of the greedy algorithm (Algorithm~\ref{algo:greedy}).}
\label{fig:greedy_illustration}
\end{figure}

 The sets $X_1$ and $X_2$ (line 4 of the algorithm) are $\scriptsize\left\{\left[\begin{array}{c} 0\\3\end{array}\right]\right\}$ and $\scriptsize\left\{\left[\begin{array}{c} 3\\0\end{array}\right]\right\}$ respectively, as also shown in Figure \ref{fig:greedy_illustration}(b). Since both sets contain a unique minimum, the algorithm  arbitrarily includes one of these two points in the sequence, that is, $ \scriptsize\left[\begin{array}{c} 3\\0\end{array}\right]$ in this case. In the next two steps, the points $\scriptsize\left[\begin{array}{c} 2\\1\end{array}\right]$ and $\scriptsize\left[\begin{array}{c} 0\\3\end{array}\right]$ are included in the sequence. In the next step illustrated in Figure~\ref{fig:greedy_illustration}(e), the sets $X_1$ and $X_2$ are $\scriptsize\left\{\left[\begin{array}{cc} 1\\2\end{array}\right],\left[\begin{array}{cc} 2\\2\end{array}\right]\right\}$ and $\scriptsize\left\{\left[\begin{array}{cc} 1\\2\end{array}\right],\left[\begin{array}{cc} 1\\3\end{array}\right]\right\}$ respectively. Since there is no unique minimum, cardinalities of $X_1$ and $X_2$ are compared (line 9 in the Algorithm~\ref{algo:greedy}) and a point from a smaller set will be chosen. In this example, $\scriptsize\left[\begin{array}{c} 2\\2\end{array}\right]$ is chosen. The sequence returned by the algorithm is as follows:

\begin{equation*}
\calD = \left[
\left[\begin{array}{c}3\\0\end{array}\right],
\left[\begin{array}{c}2\\1\end{array}\right],
\left[\begin{array}{c}0\\3\end{array}\right],
\left[\begin{array}{c}2\\2\end{array}\right],
\left[\begin{array}{c}1\\3\end{array}\right]
\right].
\end{equation*}

\bibliographystyle{IEEEtran}
\bibliography{references}

\end{document}